\documentclass[a4paper,american,apl,aip,preprintl,superscriptaddress,aip,amsmath,amssymb]{revtex4-1}
\usepackage[latin9]{inputenc}
\usepackage{color}
\usepackage{babel}
\usepackage{rotating}
\usepackage{multirow}
\usepackage{amsmath}
\usepackage{graphicx}
\usepackage[unicode=true,
 bookmarks=true,bookmarksnumbered=false,bookmarksopen=false,
 breaklinks=true,pdfborder={0 0 1},backref=false,colorlinks=true]
 {hyperref}
\hypersetup{pdftitle={Novel layered structure},
 pdfauthor={Dawei Wang},
 pdfborderstyle=}

\makeatletter

\pdfpageheight\paperheight
\pdfpagewidth\paperwidth

\providecommand{\tabularnewline}{\\}

\usepackage{babel}
\usepackage{pslatex}
\allowdisplaybreaks

\usepackage{babel}

\usepackage{bm}

\makeatother

\begin{document}
\title{ATiO$_{3}$/TiO (A=Pb, Sn) superlattice: bridging ferroelectricity
and conductivity}
\author{S. Raza}
\affiliation{School of Microelectronics \& State Key Laboratory for Mechanical
Behavior of Materials, Xi'an Jiaotong University, Xi'an 710049, China}
\affiliation{School of Energy and Environment, City University of Hong Kong, Kowloon
999077, Hong Kong Special Administrative Region, China}
\author{R. Zhang}
\affiliation{School of Engineering and Materials Science, Queen Mary University
of London, London E1 4NS, United Kindom}
\author{N. Zhang}
\affiliation{Electronic Materials Research Laboratory--Key Laboratory of the Ministry
of Education and International Center for Dielectric Research, Xi'an
Jiaotong University, Xi'an 710049, China}
\author{Z. Li}
\affiliation{School of Materials Science and Engineering, University of Science
and Technology Beijing, Beijing 100083, China}
\author{L. Liu}
\affiliation{College of Materials Science and Engineering, Guilin University of
Technology, Guilin 541004, China}
\author{F. Zhang}
\affiliation{School of Microelectronics \& State Key Laboratory for Mechanical
Behavior of Materials, Xi'an Jiaotong University, Xi'an 710049, China}
\author{D. Wang}
\email{dawei.wang@xjtu.edu.cn}

\affiliation{School of Microelectronics \& State Key Laboratory for Mechanical
Behavior of Materials, Xi'an Jiaotong University, Xi'an 710049, China}
\author{C.-L. Jia}
\affiliation{School of Microelectronics \& State Key Laboratory for Mechanical
Behavior of Materials, Xi'an Jiaotong University, Xi'an 710049, China}
\affiliation{Ernst Ruska Center for Microscopy and Spectroscopy with Electrons,
Forschungszentrum, Jülich 52425, Germany}
\date{\today}
\begin{abstract}
We propose to insert TiO layers to perovskite ATiO$_{3}$ to form
a superlattice and use first-principles calculations to investigate
its basic properties. Our computational analysis shows that the structure,
which consists of repeated ATiO$_{3}$ and TiO layers, has strong
anisotropic conductivity. The structure immediately suggests a possible
control of its conductivity by ion displacements related to its intrinsic
ferroelectricity. In addition, we have obtained the structural information
of its low-energy phases with the aid of phonon calculation and examined
their evolution with epitaxial strain. Since the number of possible
combinations is huge, we have therefore suggested an approach to mix
perovskites and simpler metal-oxides to build materials with novel
properties.
\end{abstract}
\maketitle

\section{Introduction}

The crystal structure of a material is one of the most important factors
that determines its properties. For ferroelectric and piezoelectric
materials, strain engineering has been used to enhance their performances
\citep{Helmolt_1993,Locquet_1998,Lee_2000,Bea_2009,Schlom_2014,Nakashima_2017,Zhang_2018}
by tuning the in-plane lattice constant and ion displacements, and
its influence on structural properties and polarization has been extensively
studied. In addition to applying strain, combing ferroelectricity
with other functionality (e.g., conductivity) in a single structure
can potentially induce their mutual control, crucial for technological
applications. The most important example is probably the LaAlO$_{3}$/SrTiO$_{3}$
heterointerface, where the polarity discontinuity creates high-mobility
electron gas \citep{Ohmoto2004} that enables the conductivity, which
is rather surprising since both materials are insulators. Later, Lee
\emph{et al} replaced the polar LaAlO$_{3}$ with a polarizable gel,
resulting in continuously tunable electron density on the interface
\citep{Lee2011}.

Researchers have proposed other superstructures to go beyond perovskite
(ABO$_{3}$) by combining it with other oxides. One well known combination
is a perovskite interfacing with its constituent oxide where an extra
rocksalt AO layer is added (with some in-plane shift) to $n$ layers
of ABO$_{3}$ ($n=1,2,3\dots$), resulting in the A$_{n+1}$B$_{n}$O$_{3n+1}$
structure \citep{Fennie_2005,Ruddlesden_1958} {[}see Fig.\,S1 (a)
in the supplemental material \citep{Supplemental}{]}, referred to
as the Ruddlesden-Popper (RP) phase, which has been experimentally
realized \citep{Haeni_2001,Haislmaier_2016} and its ferroelectric
nature investigated for, e.g., Sn$_{2}$TiO$_{4}$ and Pb$_{2}$TiO$_{4}$
\citep{Fennie_2005,Zhang_2014}. In addition, Aurivillius-type layred
structure has also been investigated \citep{Li_2010,Yang_2012,Yuan_2014}
and recently proposed for energy storage \citep{Tang2019}.

One may wonder if tunable conductivity can be achieved by mixing perovskite
with simpler metal-oxides. Such a possibility indeed exists by inserting
extra BO layers into a perovskite (which means TiO for ATiO$_{3}$).
While the misfit strain between ATiO$_{3}$ and TiO can be huge, given
the recent discovery of giant polarization in highly strained ($\sim16.5\%$
tensile strain) crystal structure \citep{Zhang_2018}, the ATiO$_{3}$/TiO
superlattice is plausible and worth first-principles investigation
to establish its basic properties. Furthermore, while perovskites
junctioned with TiO$_{2}$ have been explored to optimize their photoelectrochemical,
photocatylitic, and dielectric properties \citep{Jang_2017,Zeng_2018},
the use of TiO lacks any prior investigation.

\begin{figure}[h]
\begin{centering}
\includegraphics[width=8cm]{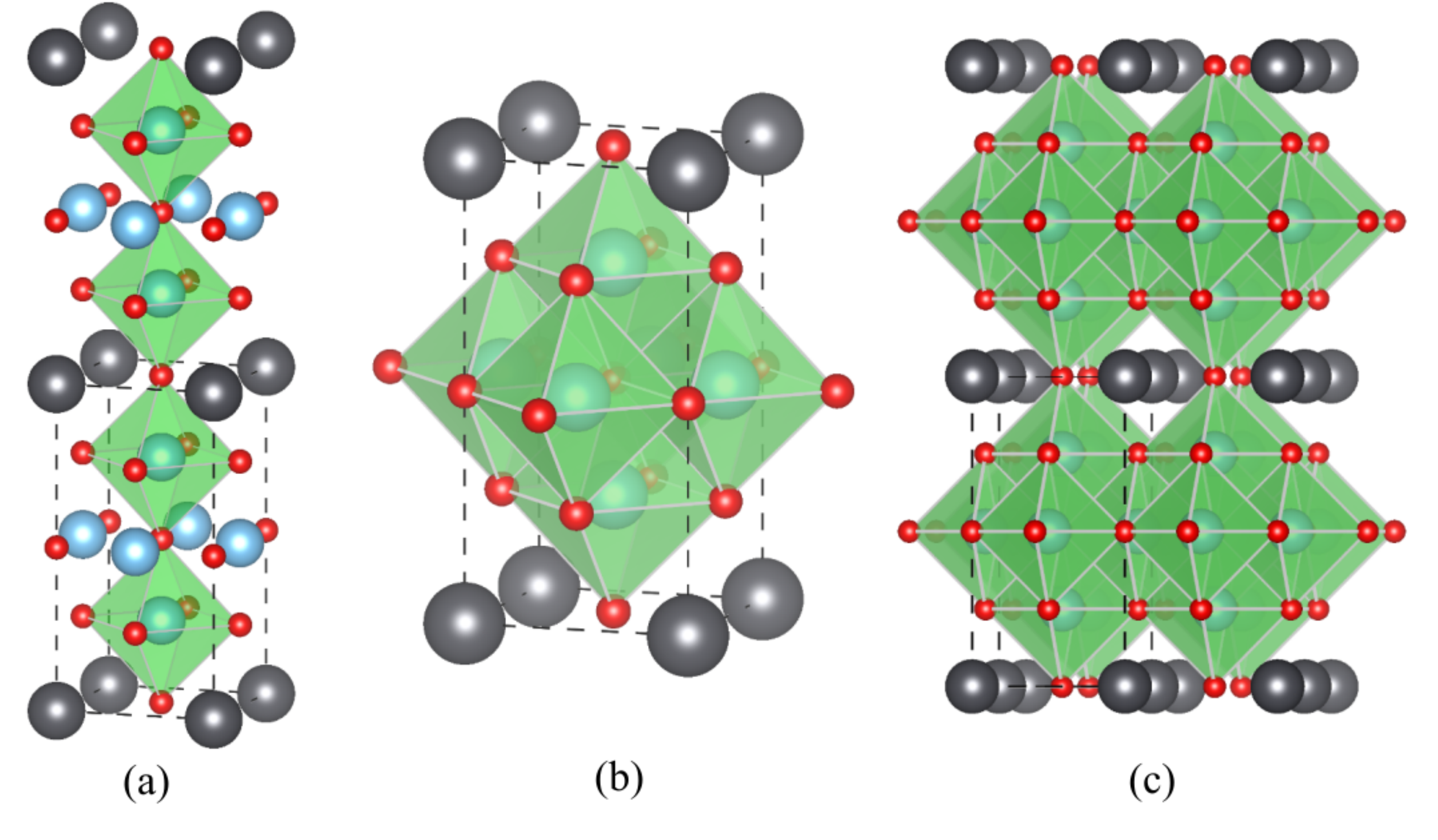}
\par\end{centering}
\caption{(a) The $1\times1\times2$ superlattice model of the proposed ATiO$_{3}$/TiO
structure, which is grown along the $z$-direction. (b-c) The $1\times1\times1$
and $2\times2\times2$ superlattice model viewed at different angles
to emphasize the oxygen octahedron inside. The dark grey, blue, and
red spheres represent A (Pb or Sn), Ti, and O atoms, respectively.\label{fig: Octahedron}}
\end{figure}
 In this work, we investigate superlattices made of ATiO$_{3}$/TiO
(A = Pb, Sn) with and without epitaxial strain. The superlattice is
built by stacking the $\left(001\right)$ planes of simple metal-oxdes,
resulting in the AO-TiO$_{2}$-$2$TiO-TiO$_{2}$ structure, which
has 12 atoms in each unit cell shown as the dotted-line-enclosed region
in Fig. \ref{fig: Octahedron} (a), noting that the TiO layer has
two Ti atoms and two O atoms in one unit cell. The general formula
for such a superlattice is (ATiO$_{3}$)$_{n}$(2TiO)$_{m}$(TiO$_{2}$)
($n=1,2,3\dots$ $m=1,3,5\dots$), different from the RP structure
(A$_{n+1}$Ti$_{n}$O$_{3n+1}$) as shown in Fig.\textcolor{red}{\,}S1
\citep{Supplemental}. The TiO layer has been properly shifted in
the $x$-$y$ plane to allow the oxygen atoms to sit above the Ti
atom of ATiO$_{3}$, maintaining the TiO$_{6}$ octahedra that strengthens
the bonding of the TiO and TiO$_{2}$ layers, where edge-sharing oxygen
octahedra can be seen in Fig. \ref{fig: Octahedron}\textcolor{red}{{}
}\textcolor{black}{(b) and (c)}. The specific arrangement of oxygen
octahedra likely deactivates their tilting. In addition to the structural
novelty of the superlattice, TiO is known to be stable and metallic
\citep{Denker1966,Ahuja1996,Ciftci2009} and nano TiO cyrstal is a
type-II superconductor \citep{AM2018} while ATiO$_{3}$ (A $=$ Pb,
Sn) is an insulator with strong ferroelectricity. Therefore, the proposed
superlattice is an ordered structure formed by repeating different
functional units, following the recently proposed research paradigm
\citep{Chen_2019}. In this paper, we will investigate their structural
and electronic properties using first-principles calculations. As
we will show, such a structure bridges ferroelectricity and conductivity
that is potentially tunable by the relative number of the TiO and
the ATiO$_{3}$ layers.

\section{Method}

In first-principles calculations, the atomic positions and lattice
constants are relaxed until both the force on an ion and the stress
fall below 0.005\,eV/Å using the projector augmented plane wave (PAW)
method as implemented in the GPAW software package \citep{Enkovaara_2010}.
The localized density approximation (LDA) is used with a cut-off energy
of 750\,eV to ensure the convergence. A $4\times4\times2$ Monkhorst-Pack
\citep{Monkhorst1976} sampling is used for the $k$-space integration.
The used valence orbitals are: Sn (5$s$ 5$p$ 4$d$), Pb (6$s$ 6$p$
5$d$), Ti (3$s$ 3$p$ 4$s$ 3$d$) and O (2$s$ 2$p$). All the
crystal structures are created using Atomic Simulation Environment
(ASE) \citep{Larsen_2017} and visualized with VESTA \citep{Vesta2011}.
With such a setup, we first obtained the tetragonal structures of
SnTiO$_{3}$ and PbTiO$_{3}$, which are compared to literature \citep{Xue_2015},
showing excellent agreements (see Tab. S1\citep{Supplemental}). The
initial lattice constants of SnTiO$_{3}$ and PbTiO$_{3}$ with ions
in their ideal positions ($P4/mmm$ symmetry) have the lattice parameters
of $a=3.79$, $c=4.175$\,Å \citep{Matar_2009}, and $a=3.89$, $c=4.15$\,Å
\citep{Zhang_2018}, respectively, whereas the rocksalt TiO has a
lattice constant of $a=4.17$\,Å \citep{Wang_2017} with the $Fm\bar{3}m$
symmetry, indicating a stretching of the perovskite part in the $x$-$y$
plane when its lattice is coherently kept with TiO. We note, under
external epitaxial strain, the lattice mismatch between ATiO$_{3}$
and TiO can be adjusted.

\section{Results and Discussion}

\begin{figure}[h]
\begin{centering}
\emph{\includegraphics[width=6cm]{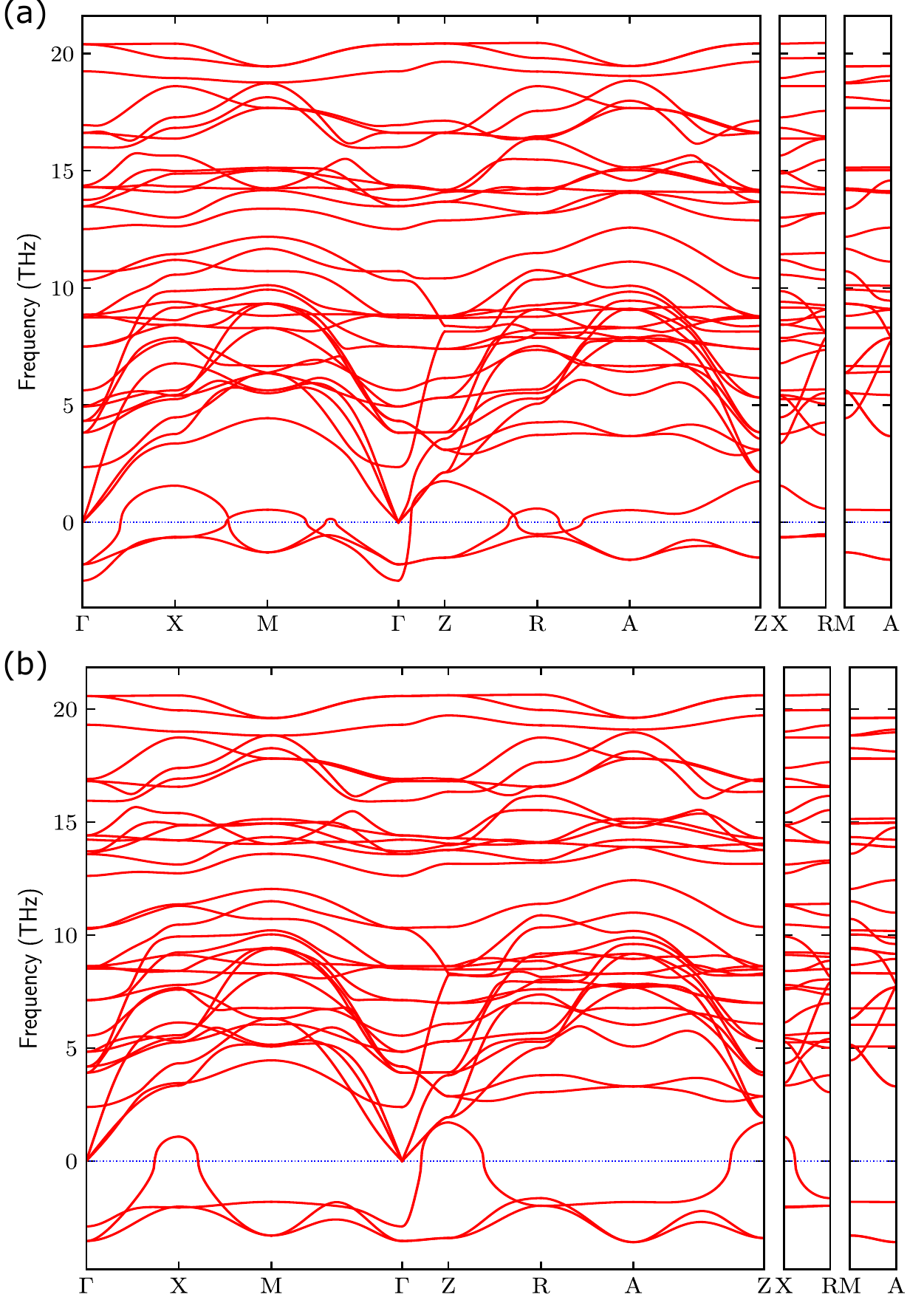}}
\par\end{centering}
\caption{Phonon band structure for PbTiO$_{3}$/TiO (a) and SnTiO$_{3}$/TiO
(b). \label{fig:Phonon-band-structure}}
\end{figure}
 We use Phonopy \citep{phonopy} to perform phonon calculation on
the original superlattice ($P4/mmm$ symmetry \citep{P4mmm}) in order
to obtain unstable phonon modes and construct low-energy structural
phases accordingly. To this end, the lattice constants of the structure
are optimized but ions are fixed on their initial (ideal) positions.
We have used a $2\times2\times2$ supercell (96 atoms) to get accurate
results on the high-symmetry points (e.g., $\Gamma$, $X$, $R$,
and $M$ points in the Brillouin-zone). Figure \ref{fig:Phonon-band-structure}
shows that the phonon dispersion has the strongest instability at
the $\Gamma$ point {[}$\boldsymbol{k}=(0,0,0)${]}, providing clues
to the most likely structural phases. We note the $\Gamma$ point
instability is inherent to many ferroelectric perovksites including
PbTiO$_{3}$ and SnTiO$_{3}$.

\begin{figure*}
\noindent \begin{centering}
\includegraphics[width=12cm]{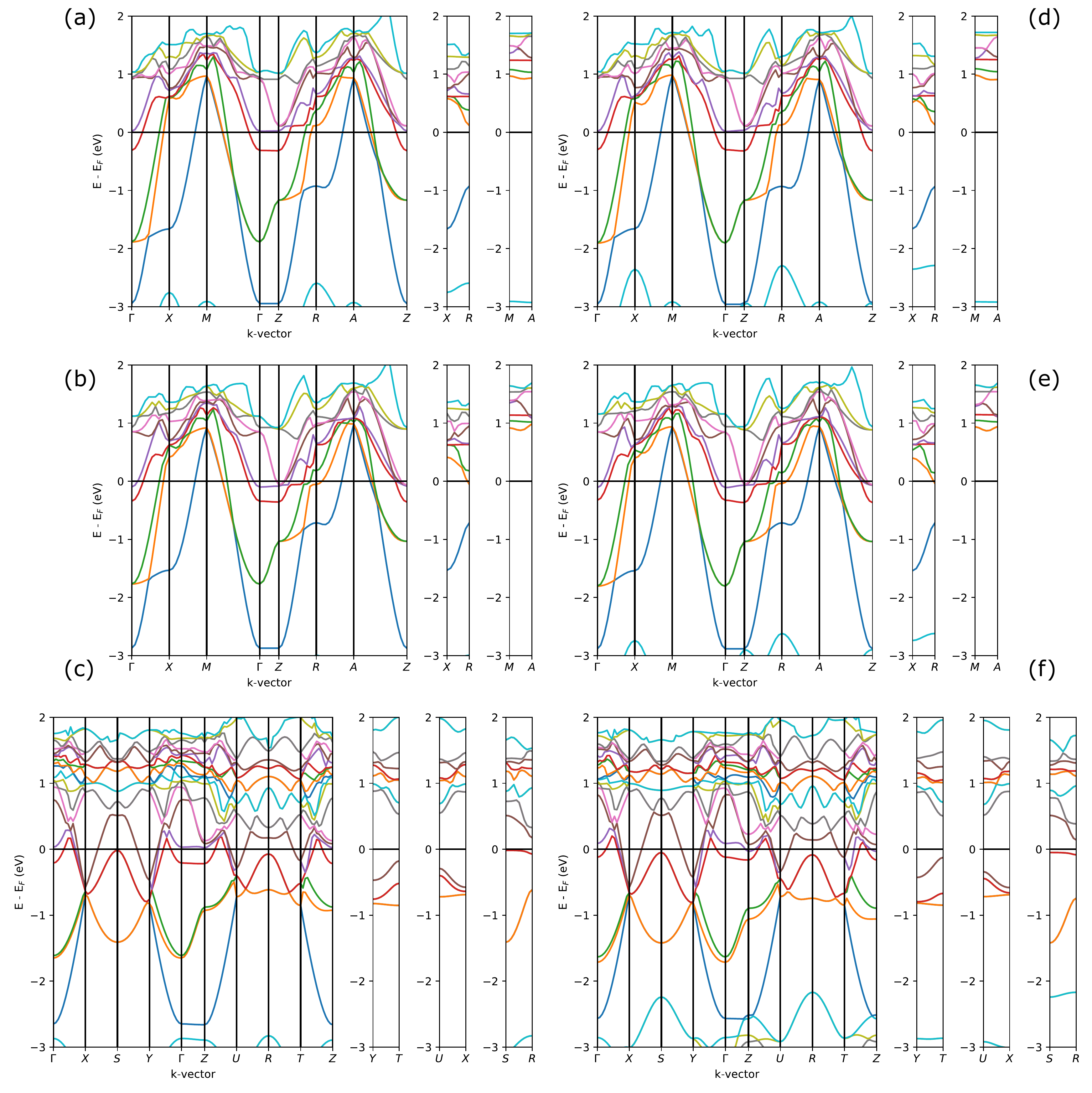}
\par\end{centering}
\caption{Bandstructure of PbTiO$_{3}$/TiO (left panel) and SnTiO$_{3}$/TiO
(right panel) in its $P4/mmm$ phase. (a) Ions in ideal positions,
cell optimized; (b) Ions and cells are both relaxed; (c) The $Amm2$
phase (using the basis of $\boldsymbol{a}+\boldsymbol{b}$ and $\boldsymbol{a}-\boldsymbol{b}$,
therefore the directions in the reciprocal space is different from
(a) and (b) ). (d-f) Counterparts of (a-c) for SnTiO$_{3}$/TiO with
(f) having the $Cm$ phase. \label{fig:bandstructure}}
\end{figure*}
\begin{table}[h]
\noindent \begin{centering}
\begin{tabular}{|c|c|c|c|c|c|c|c|c|}
\hline 
\multicolumn{2}{|c|}{Space group} & $a$(Å) & $b$(Å) & $c$ (Å) & $\alpha$ & $\beta$ & $\gamma$ & Energy\tabularnewline
\hline 
\hline 
\multirow{6}{*}{\begin{turn}{90}
SnTiO$_{3}$/TiO
\end{turn}} & $P4/mmm$ & 4.0067 & 4.0067 & 7.8791 & 90 & 90 & 90 & 172.6\tabularnewline
\cline{2-9} \cline{3-9} \cline{4-9} \cline{5-9} \cline{6-9} \cline{7-9} \cline{8-9} \cline{9-9} 
 & $P4mm$ & 4.0069 & 4.0069 & 7.9261 & 90 & 90 & 90 & 156.4\tabularnewline
\cline{2-9} \cline{3-9} \cline{4-9} \cline{5-9} \cline{6-9} \cline{7-9} \cline{8-9} \cline{9-9} 
 & $Pmm2$ & 4.0116 & 4.0117 & 7.9178 & 90 & 90 & 90 & 22.9\tabularnewline
\cline{2-9} \cline{3-9} \cline{4-9} \cline{5-9} \cline{6-9} \cline{7-9} \cline{8-9} \cline{9-9} 
 & $Amm2$ & 4.015 & 4.015 & 7.9039 & 90 & 90 & 89.33 & 22.4\tabularnewline
\cline{2-9} \cline{3-9} \cline{4-9} \cline{5-9} \cline{6-9} \cline{7-9} \cline{8-9} \cline{9-9} 
 & $Pm$ & 4.011 & 4.012 & 7.9193 & 90 & 89.98 & 90 & 22.8\tabularnewline
\cline{2-9} \cline{3-9} \cline{4-9} \cline{5-9} \cline{6-9} \cline{7-9} \cline{8-9} \cline{9-9} 
 & $Cm$ & 4.013 & 4.013 & 7.9125 & 89.98 & 89.98 & 89.54 & 0\tabularnewline
\hline 
\multirow{6}{*}{\begin{turn}{90}
PbTiO$_{3}$/TiO
\end{turn}} & $P4/mmm$ & 4.012 & 4.012 & 7.8964 & 90 & 90 & 90 & 37.4\tabularnewline
\cline{2-9} \cline{3-9} \cline{4-9} \cline{5-9} \cline{6-9} \cline{7-9} \cline{8-9} \cline{9-9} 
 & $P4mm$ & 4.013 & 4.013 & 7.8974 & 90 & 90 & 90 & 37.3\tabularnewline
\cline{2-9} \cline{3-9} \cline{4-9} \cline{5-9} \cline{6-9} \cline{7-9} \cline{8-9} \cline{9-9} 
 & $Pmm2$ & 4.023 & 4.000 & 7.8946 & 90 & 90 & 90 & 5.6\tabularnewline
\cline{2-9} \cline{3-9} \cline{4-9} \cline{5-9} \cline{6-9} \cline{7-9} \cline{8-9} \cline{9-9} 
 & $Amm2$ & 4.012 & 4.012 & 7.8950 & 90 & 90 & 89.88 & 0\tabularnewline
\cline{2-9} \cline{3-9} \cline{4-9} \cline{5-9} \cline{6-9} \cline{7-9} \cline{8-9} \cline{9-9} 
 & $Pm$ & 4.023 & 4.000 & 7.8957 & 90 & 89.99 & 90 & 5.4\tabularnewline
\cline{2-9} \cline{3-9} \cline{4-9} \cline{5-9} \cline{6-9} \cline{7-9} \cline{8-9} \cline{9-9} 
 & $Cm$ & 4.012 & 4.012 & 7.8946 & 89.98 & 89.98 & 89.87 & 0.1\tabularnewline
\hline 
\end{tabular}
\par\end{centering}
\caption{Structural information of the ATiO$_{3}$-TiO (A=Sn,Pb) superlattice
after relaxation. Note the rocksalt TiO has a lattice constant of
4.177\,Å and the energy unit is meV per formula.\label{tab:Properties-of-different-1}}
\end{table}
 Based on the phonon spectrum and common ferroelectric phases of perovskites,
we obtained several structural phases of the ATiO$_{3}$/TiO superlattice
by displacing Ti in the perovskite layer along the high-symmetry directions,
forming the $P4/mmm,P4mm,Pm,Amm2,Pmm2$, and $Cm$ phases \citep{Dieguez_2008}.
The overall symmetry of the superlattice is checked both before and
after their structural relaxation. Table \ref{tab:Properties-of-different-1}
displays the information of the relaxed SnTiO$_{3}$/TiO and PbTiO$_{3}$/TiO,
which have the $Cm$ and $Amm2$ phase as the ground state, respectively.
The perovskite part of the relaxed ATiO$_{3}$/TiO experiences a tensile
strain of about $3$-$5\%$ while the TiO part undergoes a compression
of about $4\%$. The tensile strain forces the ATiO$_{3}$ component
into a pseudocubic structure ($a=b\simeq c/2$). The in-plane lattice
constants of SnTiO$_{3}$/TiO and PbTiO$_{3}$/TiO are rather close
since the rigid TiO layer is unable to compress further. Except the
$P4mm$ and $P4/mmm$ phases, the structural phases have close energies,
indicating that the ion displacement inside the $x$-$y$ plane is
preferred. In addition, the SnTiO$_{3}$/TiO has a stronger tendency
to stay in the $Cm$ phase with a nonzero polarization along $z$,
an inherent property to SnTiO$_{3}$\citep{Wang_2018}.

Table S2 \citep{Supplemental} provides the ion displacements (in
fractional coordinates) for the $Cm$ and $Amm2$ phases, showing
large displacements of ions in the $x-y$ plane and relatively small
shifts in the $z$ direction. It also indicates that SnTiO$_{3}$/TiO
has larger ion displacements than PbTiO$_{3}$/TiO. Due to its smaller
ionic size, the Sn atom exhibits a much larger (10 times larger) displacement
than the Pb atoms. To further confirm that the $Cm$ ($Amm2$) phase
is the ground state, we have also constructed other configurations
suggested by the $\Gamma$ point phonon eigenvector, removed all the
symmetry constraints, and relaxed the structure again; but no new
structural phases with lower energy were found.

Let us now turn to the electronic properties of ATiO$_{3}$/TiO. Since
PbTiO$_{3}$ and SnTiO$_{3}$ are insulators, it is important to know
how the TiO layer alters the electronic band structures. For TiO,
we have calculated its band structure, agreeing well with literature
\citep{Denker1966,Ahuja1996,Ciftci2009}, confirming that TiO is a
conductor (see Fig.\,S2 \citep{Supplemental}). Figure \ref{fig:bandstructure}
displays the electronic band structure of PbTiO$_{3}$/TiO (left column)
and SnTiO$_{3}$/TiO (right column). Figure \ref{fig:bandstructure}(a)
shows that the electronic band structure of PbTiO$_{3}$/TiO in its
original $P4/mmm$ phase, which clearly demonstrates that the superlattice
is still a conductor, while the $\Gamma$ to $Z$ line (related to
the superlattice growth direction) is slightly above the Fermi level.
Figure \ref{fig:bandstructure}(b) shows the band structure when the
ions are also relaxed. This structure is still a conductor, but the
$\Gamma$ to $Z$ line has dropped below the Fermi level. While such
a change is small, it implies a strong influence on the conductivity
along the growth direction. For the lowest energy state we have found
(with the $Amm2$ symmetry), Fig.\,\ref{fig:bandstructure}(c) shows
that the $\Gamma$ to $Z$ line again rises above the Fermi level,
but remains very close to it.

The electronic band structure along the $\Gamma$-$Z$ line shown
in Figs.\,\ref{fig:bandstructure}(a-c) indicates that the conductivity
along the superlattice-growth direction can be quite sensitive to
ion displacements. This connection is potentially important given
the fact that both PbTiO$_{3}$ and SnTiO$_{3}$ are ferroelectric,
a feature arising from their ion displacements. Moreover, in all the
three cases shown in Figs.\,\ref{fig:bandstructure}(a-c), the $\Gamma$
to $Z$ line is special in its flatness, indicating a huge electron/hole
effective mass along the superlattice-growth direction, resulting
in strong anisotropic conductivity. Figures \ref{fig:bandstructure}(d-f)
show that SnTiO$_{3}$/TiO has similar behaviors on the $\Gamma$
to $Z$ line with some minor variations comparing to PbTiO$_{3}$.
Similar behavisor with the Ruddlesden-Popper double-layered perovskite
La$_{3}$Ni$_{2}$O$_{7}$ has been reported in recent years \citep{Lee2014,Mochizuki2018}.

\begin{figure}[h]
\begin{centering}
\includegraphics[width=8cm]{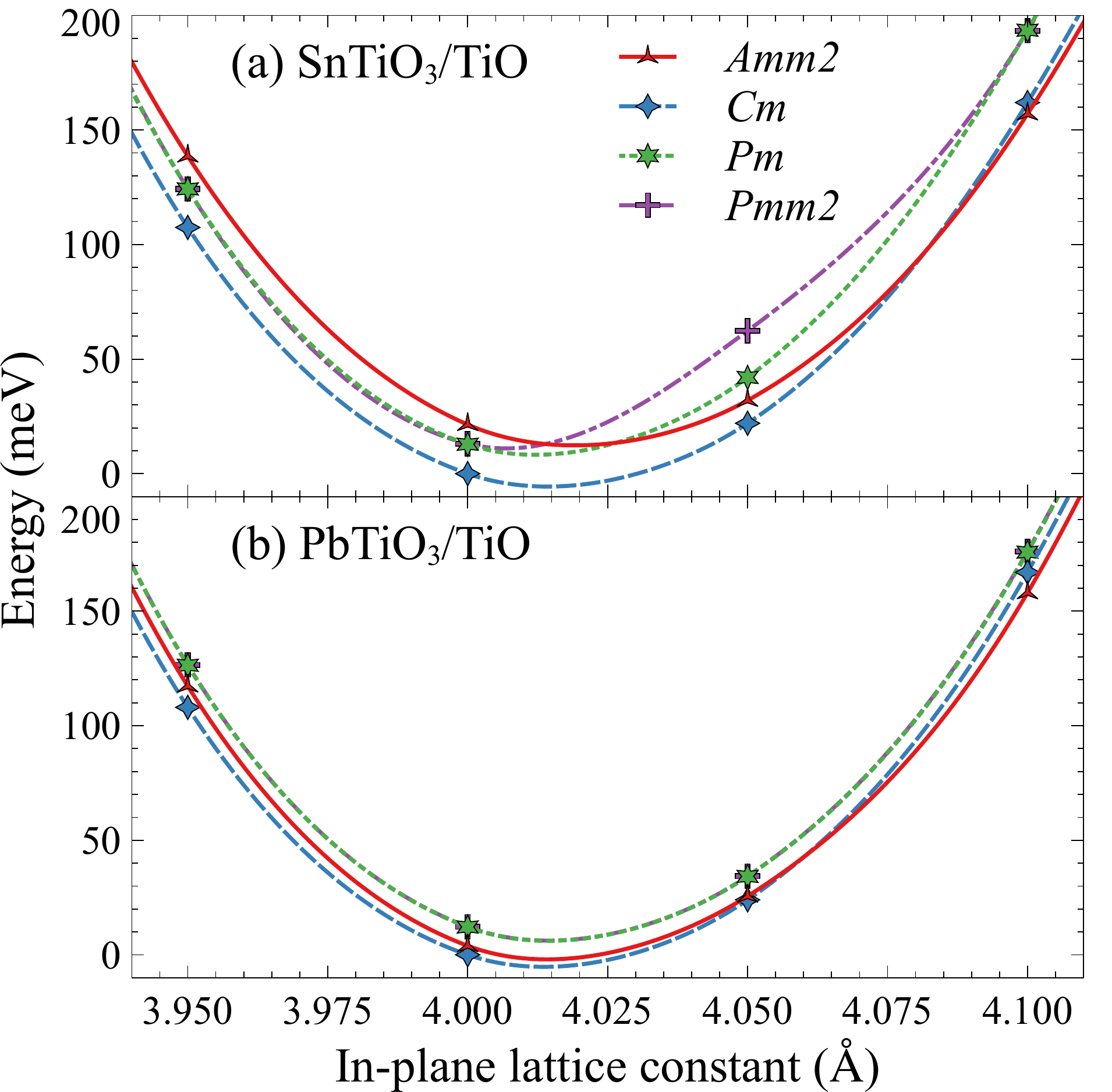}
\par\end{centering}
\caption{The energy versus the lattice constant for epitaxially strained SnTiO$_{3}$/TiO
(a) and PbTiO$_{3}$/TiO (b) superlattice with the $Amm2$, $Pmm2,Pm,$and
$Cm$ symmetries.\label{fig:ATiO-strain-1}.}
\end{figure}
 Finally, let us briefly discuss epitaxial strain effects on ATiO$_{3}$/TiO.
We replicate the biaxial epitaxial strain by fixing the in-plane lattice
constant. The misfit strain is calculated with $\varepsilon=\left(a-a_{0}\right)/a_{0}$
where $a_{0}$ is the optimized lattice constant of the superlattice
without the in-plane constraint. The energies of the structural phases
versus the epitaxial strain are shown in Fig.\,\ref{fig:ATiO-strain-1},
where the $P4mm$ and $P4/mmm$ phases are omitted due to their much
higher energy. Figure \ref{fig:ATiO-strain-1}(a) shows that SnTiO$_{3}$/TiO
has the $Cm$ phase as the ground state for a wide range of mistfit
strain until $a=4.10$\,Å ($\simeq2.15\%$ tensile strain) when the
$Amm2$ phase becomes most stable. Figure \ref{fig:ATiO-strain-1}(b)
shows a similar behavior for the PbTiO$_{3}$/TiO where the $Cm$
is most stable phase until $a=4.05$\,Å ($\simeq0.95\%$ tensile
strain) \citep{Cm_is_lower}. Over the whole misfit strain range,
the $Pm$ {[}with polarization direction $(u,0,w)$ {]} and $Pmm2$
{[}$(u,0,0)${]} phases are close in energy to the $Amm2$ {[}$(u,u,0)${]}
and the $Cm$ {[}$(u,u,w)${]} phases while their ion displacements
are in different directions. Figure S3 shows the $\left(c/2\right)/a$
ratio for ATiO$_{3}$/TiO, which is adopted to compare to the $c/a$
of simple ATiO$_{3}$, decreases with the in-plane lattice constant.
The $\left(c/2\right)/a$ of SnTiO$_{3}$/TiO remains higher than
that of PbTiO$_{3}$/TiO since the tetragonality of the parent SnTiO$_{3}$($1.13)$
\citep{Parker_2011} is larger than PbTiO$_{3}$($1.065)$ \citep{Saghi-Szabo_1998}. 

The proposed superlattice is composed of perovskite ATiO$_{3}$ and
TiO layers. As a matter of fact, there are other layered metal-oxides
in addition to the RP phase \citep{cava_page}. The Aurivillius phase
was first proposed by Aurivillius with the general formula (Bi$_{2}$O$_{2}$)(A$_{n-1}$B$_{n}$O$_{3n+1}$),
where A and B are cations \citep{Kendall1996}. It can be thought
of as alternating layers of Bi$_{2}$O$_{2}$ and perovskites whose
ferroelectric, magnetic, and dielectric properties have been extensively
studied in the past \citep{Yang_2012,Yuan_2014,Li_2010}. The Dion--Jacobson
series has the general formula A$^{\prime}$ (A$_{n-1}$B$_{n}$O$_{3n+1}$),
where A$^{\prime}$, A and B are cations \citep{Choy2001,Fennie_2006}.
This structure has an intermediate layer of A$^{\prime}$ atoms (often
alkali metal) between perovskite blocks \citep{Fennie_2006}. More
recently, Tsujimoto \emph{et al} managed to remove more oxygens from
conventional perovskite SrFeO$_{3-x}$, producing a structure with
alternating layers of FeO$_{2}$ and Sr \citep{Tsujimoto2007}. The
superlattice we proposed here are different from these known layered
structures.

There are many simple oxides and perovskites to choose from to build
the proposed superlattice. For instance, possible monoxides with the
rocksalt structures include MgO ($a=4.217$\,Å), MnO ($a=4.446$\,Å),
NiO ($a=4.178$\,Å), CoO ($a=4.263$\,Å), and NbO ($a=4.2101$ Å).
Among them, NbO is metallic \citep{Tao2011,Dhamdhere_2016} and NiO
and CoO are antiferromagnetic. Such combinations have the potential
to generate novel materials with special properties. Since the growth
of the RP phase has already used techniques like reactive/hybrid molecular
beam epitaxy \citep{Haeni_2001,Haislmaier_2016}, the growth of the
proposed superlattice does not seem particularly difficult, also noting
that TiO does not add any new element to ATiO$_{3}$. The ATiO$_{3}$/TiO
superlattice can be fabricated by alternating growth of the ATiO$_{3}$
and TiO layers using oxide molecular beam epitaxy (O-MBE) \citep{Mochizuki2018},
pulsed laser deposition (PLD), or atomic layer deposition (ALD) methods.
Monitoring with the \emph{in situ} reflected high energy electron
diffraction (RHEED), the thickness of both the ABO$_{3}$ and the
TiO layers, as well as their periodicity, can be precisely controlled.
The epitaxial strain can be induced by properly selecting single crystal
perovskite substrates. For instance, the BaTiO$_{3}$ (001) and $(1-x)$PbMg$_{1/3}$Nb$_{2/3}$O$_{3}$-$x$PbTiO$_{3}$
substrates can induce tensile strain while the SrTiO$_{3}$ (001),
LaAlO$_{3}$ (001) substrates can induce compressive strain.

\section{Conclusion}

In summary, we have proposed a superlattice with TiO layers inserted
into the perovskite PbTiO$_{3}$ or SnTiO$_{3}$, investigated their
structural and electronic properties, finding that they are anisotropic
conductors with tunable conductivity that is sensitive to ion displacements.
Such a feature can couple conductivity to ferroelectricity, enabling
the control of conductivity with electric field or strain. Furthermore,
the proposed structure is automatically a micro-multilayer ceramic
capacitor and can have large permittivity due to the Maxwell-Wagner
effect. We finally note that, in growing such superlattice, there
is no constraint on the ratio between the number of ATiO$_{3}$ layers
and the TiO layers, which can be used to tune special properties of
the resulting superlattices.
\begin{acknowledgments}
This work is financially supported by the National Natural Science
Foundation of China, Grant No. 11974268, 11574246, and U1537210. D.W.
also thanks support from the Chinese Scholarship Council (201706285020).
\end{acknowledgments}


\begin{thebibliography}{99}
\bibitem{P4mmm} We note that alternately stacking ATiO$_{3}$ layer
with TiO layer is equivalent to replacing every second AO layer with
the TiO layer, resulting in the $P4/mmm$ symmetry if the original
cubic perovksite cells are not distorted.
\bibitem{Cm_is_lower} For epitaxially strained PbTiO$_{3}$/TiO,
the results here indicate that the $Cm$ phase can be slightly lower
in energy (\textasciitilde 0.16\,meV) than the $Amm2$ phase, different
from the result shown in Tab. \ref{tab:Properties-of-different-1}.
This is because the angle between the in-plane $x$ and $y$ axis
($\gamma)$ is not allowed to change when epitaxially strained, making
it different from the free relaxation shown in Tab. \ref{tab:Properties-of-different-1}.
\end{thebibliography}
\end{document}